\documentclass[preprint, aps, showpacs, superscriptaddress, floatfix, longbibliography]{revtex4-1}
\usepackage{graphicx}
\graphicspath{.,{../figures/}}
\usepackage[usenames,dvipsnames]{color}
\usepackage[bookmarks=true]{hyperref}

\newcommand{\kaos}{{\sc Kaos}\@}
\newcommand{\HH}{$^4_\Lambda$H\@}

\begin{document}
%%%%%%%%%%%%%%%%%%%%%%%%%%%%%%%%%%%%%%%%%%%%%%%%%%%%%%%%%%%%%%%%%%%%%%%%%%%%%%%

% -------------------------------------------------------------------
\title{Observation of $_{\Lambda}^{4}$H hyperhydrogen by decay-pion
  spectroscopy in electron scattering}
% -------------------------------------------------------------------

\newcommand{\kph}{\affiliation{
    Institut f\"{u}r Kernphysik,
    Johannes Gutenberg-Universit\"{a}t, 
    D-55099 Mainz, Germany}}
\newcommand{\gsi}{\affiliation{
    GSI Helmholtz Centre for Heavy Ion Research, 
    D-64291 Darmstadt, Germany}}
\newcommand{\hr}{\affiliation{
    Department of Physics, 
    University of Zagreb, 
    HR-10002 Zagreb, Croatia}}
\newcommand{\si}{\affiliation{
    Department of Physics, 
    University of Ljubljana, 
    and Jo\v{z}ef Stefan Institute, 
    SI-1000 Ljubljana, Slovenia}}
\newcommand{\jp}{\affiliation{
    Department of Physics, 
    Tohoku University, 
    Sendai, 980-8571, Japan}}
\newcommand{\am}{\affiliation{
    Yerevan Physics Institute, 
    375036 Yerevan, Armenia}}
\newcommand{\him}{\affiliation{
    Helmholtz Institute Mainz, 
    D-55099 Mainz, Germany}}
\newcommand{\fiu}{\affiliation{
    Department of Physics, 
    Florida International University,
    Miami, Florida 33199, USA}}
\newcommand{\va}{\affiliation{
    Department of Physics, 
    Hampton University, 
    Hampton, Virginia 23668, USA}}

\author{A.~Esser}\kph
\author{S.~Nagao}\jp
\author{F.~Schulz}\kph
\author{P.~Achenbach} 
\email[]{patrick@kph.uni-mainz.de}\kph%\homepage{http://www.kph.uni-mainz.de}
\author{C.~Ayerbe Gayoso}\kph
\author{R.~B\"ohm}\kph
\author{O.~Borodina}\kph\gsi
\author{D.~Bosnar}\hr
\author{V.~Bozkurt}\gsi
\author{L.~Debenjak}\si
\author{M.\,O.~Distler}\kph
\author{I.~Fri\v{s}\v{c}i\'c}\hr
\author{Y.~Fujii}\jp
\author{T.~Gogami}\thanks{Present address: Graduate School of Science, Kyoto University, Kyoto 606-8502, Japan.}\jp
\author{O.~Hashimoto}\thanks{deceased}\jp
\author{S.~Hirose}\jp
\author{H.~Kanda}\jp
\author{M.~Kaneta}\jp
\author{E.~Kim}\gsi
\author{Y.~Kohl}\kph
\author{J.~Kusaka}\jp
\author{A.~Margaryan}\am
\author{H.~Merkel}\kph
\author{M.~Mihovilovi\v{c}}\kph
\author{U.~M\"uller}\kph 
\author{S.~N.~Nakamura}\jp
\author{J.~Pochodzalla}\kph
\author{C.~Rappold}\gsi\him
\author{J.~Reinhold}\fiu
\author{T.\,R.~Saito}\kph\gsi\him
\author{A.~Sanchez Lorente}\him 
\author{S.~{S\'anchez Majos}}\kph
\author{B.\,S.~Schlimme}\kph
\author{M.~Schoth}\kph
\author{C.~Sfienti}\kph
\author{S.~\v{S}irca}\si
\author{L.~Tang}\va
\author{M.~Thiel}\kph
\author{K.~Tsukada}\thanks{Present address: Research Center 
  for Electron Photon Science, Tohoku University, Sendai 982-0826, 
  Japan.}\jp
\author{A.~Weber}\kph
\author{K.~Yoshida}\gsi

\collaboration{A1 Collaboration}\noaffiliation

\date{\today}

\begin{abstract}
  At the Mainz Microtron MAMI, the first high-resolution pion
  spectroscopy from decays of strange systems was performed by
  electron scattering off a $^9$Be target in order to study the
  $\Lambda$ binding energy of light hypernuclei. Positively charged
  kaons were detected by a short-orbit spectrometer with a broad
  momentum acceptance at zero degree forward angles with respect to
  the beam, efficiently tagging the production of strangeness in the
  target nucleus.  In coincidence, negatively charged decay-pions were
  detected by two independent high-resolution spectrometers.  About
  $10^3$ pionic weak decays of hyperfragments and hyperons were
  observed.  The pion momentum distribution shows a monochromatic peak
  at $p_\pi\approx$ 133\,MeV$\!/c$, corresponding to the unique
  signature for the two-body decay of hyperhydrogen
  $^4_\Lambda\mathrm{H}\rightarrow \mathrm{^4He} + \pi^-$, stopped
  inside the target. Its $\Lambda$ binding energy was determined to be
  $B_\Lambda = 2.12 \pm 0.01\ \mathrm{(stat.)} \pm
  0.09\ \mathrm{(syst.)}$\,MeV with respect to the $^3\mathrm{H} +
  \Lambda$ mass.
\end{abstract}

 % Charged-particle spectroscopy, 29.30.Ep

\maketitle

%%%%%%%%%%%%%%%%%%%%%%%%%%%%%%%%%%%%%%%%%%%%%%%%%%%%%%%%%%%%%%%%%%%%%%%%%%%%%%%
\paragraph{Introduction---\hspace{-3mm}}
%%%%%%%%%%%%%%%%%%%%%%%%%%%%%%%%%%%%%%%%%%%%%%%%%%%%%%%%%%%%%%%%%%%%%%%%%%%%%%%

% ---- hypernuclei ----

%% A very interesting phenomenon in nuclear physics is the existence of
%% nuclei containing $\Lambda$ hyperons. 
When a $\Lambda$ hyperon replaces one of the nucleons ($N = n$ or $p$)
in the nucleus, a bound system can be formed by the hyperon and the
core of the remaining nucleons, $\Lambda$ hypernucleus.  The
\HH\ nucleus is a heavy isotope of the element hydrogen, in which a
$\Lambda$ hyperon is bound to a tritium core. It was found in early
helium bubble chamber~\cite{Block1964} and nuclear emulsion
experiments~\cite{Gajewski1967,Bohm1968,Juric1973,Bertrand1970}.  In a
ground-state, a hypernucleus decays to a non-strange nucleus through
mesonic (MWD) or non-mesonic (NMWD) weak decay modes.  By detecting
the decay of hypernuclei and measuring the momenta of the decay
products, the binding energies of the $\Lambda$ hyperon, i.e.\ the
$\Lambda$ separation energy, for a larger number of $s$- and $p$-shell
hypernuclei were reported in the 1960s and 1970s.

% ---- NN and YN interactions ----

Precise determination of the binding energies of hypernuclei can be
used to test the $YN$ interactions ($Y = \Lambda$ or $\Sigma$) in
many-body systems. Contrary to the non-strange sector, where a large
data base is used to successfully model the $NN$ forces, the available
data on $YN$ scattering is not sufficient to determine realistic
interactions among hyperons and nucleons.  Various hypernuclear
structure theories exist in which the binding energies of light
hypernuclei are calculated, most recent approaches include cluster
models and {\em ab initio} calculations with the interactions
constructed either in the meson-exchange picture or within chiral
effective field
theory~\cite{Hiyama2001:PRC65,Nogga2002,Nemura2002,Hiyama2009:PRC80,Wirth2014,Haidenbauer2007}.
Measuring the binding energy splitting in the mass $A = 4$
hypernuclei, $^{4}_\Lambda$H and $^{4}_\Lambda$He, is especially
helpful for investigating the origin of charge symmetry breaking in
the strong interaction~\cite{Nogga2013,Gal2015}.

The most abundant decay of \HH\ observed in nuclear emulsions is the
charged two-body mode $^4_\Lambda\mathrm{H}\rightarrow \mathrm{^4He} +
\pi^-$. However, it could not be used to deduce the binding energy because
of the larger systematic error in the pion range-energy relation for
pion ranges greater than 3\,cm~\cite{Juric1973}. Instead, 
only $\sim 160$ three-body decays 
%compiled from several experiments 
were used. Refs.~\cite{Gajewski1967,Bohm1968,Juric1973} evaluated 21,
63, and 56 events from $^4_{\Lambda}\mathrm{H} \rightarrow \pi^- +
\mathrm{^{1}H} + \mathrm{^{3}H}$ and only 2, 7, and 11 events from
$^4_{\Lambda}\mathrm{H} \rightarrow \pi^- + \mathrm{^{2}H} +
\mathrm{^{2}H}$, respectively. In Ref.~\cite{Juric1973} the binding
energies of decays from these two decay modes were reported separately
to be $1.92 \pm 0.12$\,MeV and $2.14 \pm 0.07$\,MeV with a difference
of $0.22 \pm 0.14$\,MeV. The FWHM of the distribution of \HH\ binding
energies is 2.1\,MeV.
%% The statistical situation is better for
%% $^4_{\Lambda}\mathrm{He}$, where a total of 279 events were analyzed
%% and the FWHM of the distribution is 1.2\,MeV. 
Despite extensive calibrations, systematic uncertainties of at least
0.05\,MeV in the binding energies should be assumed for emulsion
data~\cite{Davis1992}.  In Ref.~\cite{Gajewski1967} a possible
systematic error of 0.15\,MeV is quoted. From these ambiguities it is
evident that independent, high-resolution experiments are needed to
confirm the emulsion data on binding energies.

%% Binding energy differences in heavier mirror hypernuclei show an
%% unexplained trend of perhaps even flipping sign with respect to the $A
%% = 4$ difference, as demonstrated for $A = 7$ in a recent Jefferson Lab
%% experiment~\cite{Nakamura2013}; see also Table~3 of
%% Ref.~\cite{Davis1992}. A coherent theoretical description based on
%% realistic $Y N$ interactions or phenomenological $\Lambda N$ CSB
%% interactions is not yet found.  Therefore, the precise level and the
%% origin of CSB in the $\Lambda N$ interactions are important issues
%% related to our understanding of the fundamental baryon-baryon
%% interactions.

This Letter presents the first result of the measurement of a
$\Lambda$ binding energy in light hypernuclei from pionic decays in
electron scattering. Pions were detected in high-resolution
spectrometers measuring the momentum with much better resolution than
with emulsion and with completely independent systematical
uncertainties.

%%%%%%%%%%%%%%%%%%%%%%%%%%%%%%%%%%%%%%%%%%%%%%%%%%%%%%%%%%%%%%%%%%%%%%%%%%%%%%%
\paragraph{Measurement technique---\hspace{-3mm}}
%%%%%%%%%%%%%%%%%%%%%%%%%%%%%%%%%%%%%%%%%%%%%%%%%%%%%%%%%%%%%%%%%%%%%%%%%%%%%%%

% ----- hypernuclear reaction spectroscopy ----

In reaction spectroscopy, ground- and excited hypernuclear states can
be identified by a missing mass analysis of the incident beam and the
associated reaction meson. Since these reactions require stable target
nuclei, hypernuclei accessible by these reactions are limited. The
direct reaction spectroscopy of \HH\ is not possible using charged
meson beams in the established $(\pi^+,K^+)$ and $(K^-,\pi^-)$
reactions. The first observation of \HH\ bound states in missing mass
spectroscopy using the $(e,e'K^+)$ reaction was reported a decade
ago~\cite{Dohrmann2004}. This pioneering experiment reached a mass
resolution of $4$\,MeV$\!/c^2$. Precision measurements of $\Lambda$
binding energies were conducted at Jefferson Lab, recently the
spectroscopy of $^{7}_{\Lambda}$He with a binding energy resolution of
$\sim 0.6$\,MeV~\cite{Nakamura2013}.

% ----- method of decay-pion spectroscopy ----

The first counter experiments detecting the two-body MWD of \HH\ were
performed at KEK in the 1980s and 90s~\cite{Tamura1989,Outa1998}. From
the observed decay-pion peak at $132.6 \pm 0.3$\,MeV$\!/c$ a binding
energy of $2.35 \pm 0.22$\,MeV can be calculated. In this experiment
strangeness was exchanged with nuclei by $K^-$ absorption within a
thick target so that the momentum resolution of $1.9-3.3$\,MeV$\!/c$
FWHM was not competitive to the emulsion data in determining the
$\Lambda$ binding energy.
% the spectrometer had a low efficiency below $p_\pi \sim$
% 120\,MeV$\!/c$.

In 2007 the usage of magnetic spectrometers to measure the momenta of
pions from two-body decays of light hypernuclei fragmented from the
excited states of initially electro-produced hypernuclei was proposed
for Jefferson Lab~\cite{Tang2007}. 
%With large momentum transfer, 
An electro-produced hypernucleus can have excitation energies above
the lowest particle emission threshold and then loose excitation
energy through fragmentation, i.e.\ nucleon or cluster emission. This
is a very fast process that can lead to particle-stable hypernuclei in
a large range of mass and atomic numbers, including hyperisotopes
which are not accessible in missing-mass experiments. MWD takes place
from the ground-state of these hypernuclei. The pion momenta from
two-body MWD of the lightest systems ($A \le 9$) are $p_\pi \sim
96-138$\,MeV$\!/c$. The mass of a hypernucleus can be obtained from a
measurement of $p_\pi$. Kaons can be tagged to suppress non-strange
processes.

%%%%%%%%%%%%%%%%%%%%%%%%%%%%%%%%%%%%%%%%%%%%%%%%%%%%%%%%%%%%%%%%%%%%%%%%%%%%%%%
\paragraph{Experiment---\hspace{-3mm}}
%%%%%%%%%%%%%%%%%%%%%%%%%%%%%%%%%%%%%%%%%%%%%%%%%%%%%%%%%%%%%%%%%%%%%%%%%%%%%%%

% ----- experimental setup ----

%
% -------------------------------------------------------------------
\begin{figure}
  \centering
  \includegraphics[angle=0,width=\columnwidth]{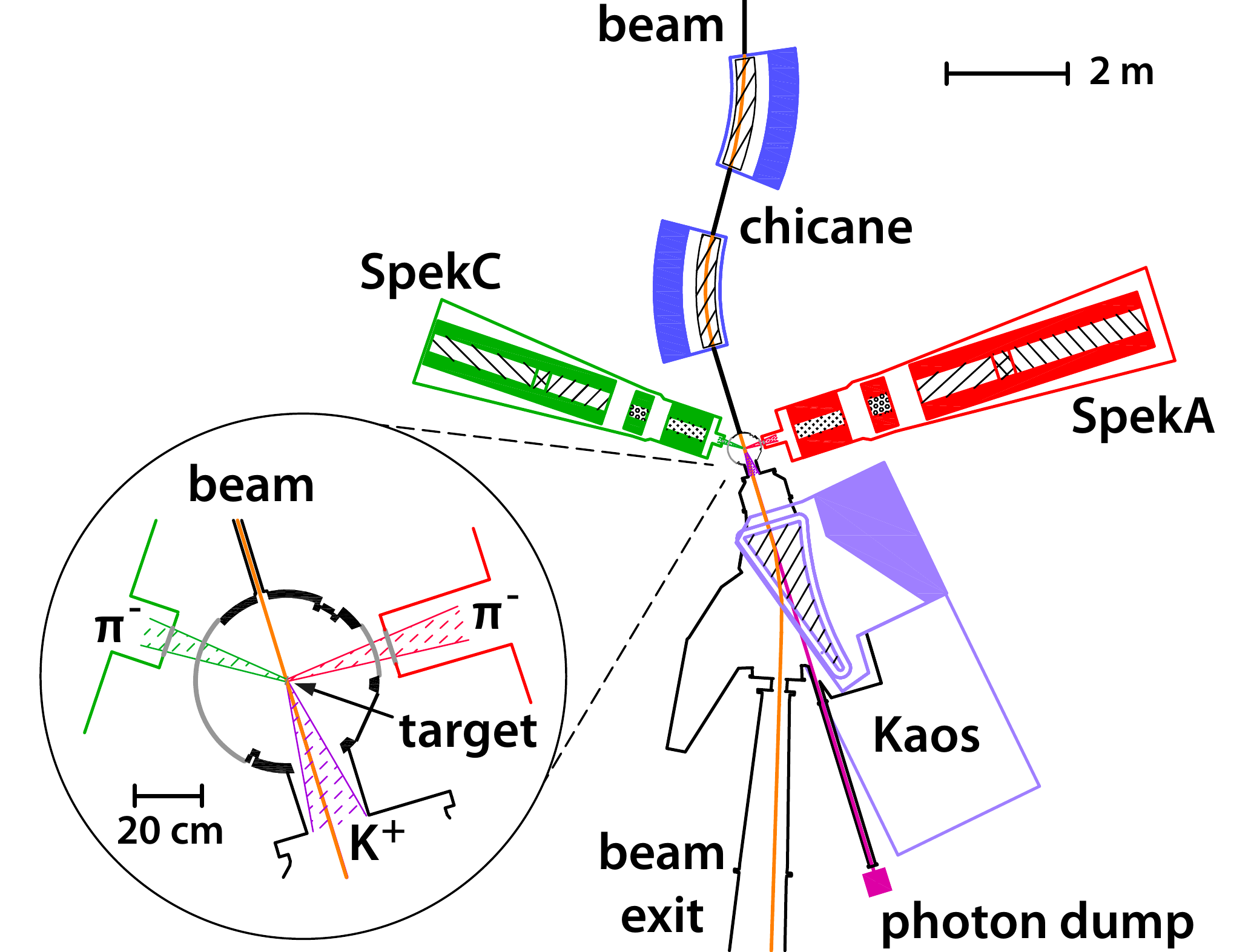}
  \caption{(color online). Layout of the setup showing the electron
    beam-line, the two high-resolution spectrometers SpekA and SpekC
    used for pion detection in the backward hemisphere, and the
    \kaos\ spectrometer at zero degree angle relative to the outgoing
    beam used for kaon tagging. The beam enters from the top. The
    inset shows an enlarged view of the scattering chamber and the
    detected particle's names.}
  \label{fig:Setup}
\end{figure}
% -------------------------------------------------------------------
%

The experiment was carried out by the A1 Collaboration at the
spectrometer facility (see Refs.~\cite{Achenbach2011,Blomqvist1998}
for a detailed description of the spectrometers) at the Mainz
Microtron MAMI-C,
%~\cite{Kaiser2008}
in Germany, with a $1.508\,$GeV electron beam incident on a
125\,$\mu$m thick and 54$^\circ$ tilted $^9$Be target foil with a beam
current of $20\,\mu$A. The layout of the experimental setup is shown
in Fig.~\ref{fig:Setup}, detail can be found in Ref.~\cite{Esser2013}.
The luminosity corrected for the data acquisition dead-time was
$\int\! {\cal L}dt \sim$ 235\,fb$^{-1}$, integrated over a period of
$\sim 250$\,h during the year of 2012. The total charge of the
electrons hitting the target was 18.2\,C.

% ----- pion detection ----

Pions were detected with two high-resolution spectrometers (SpekA and
SpekC) with quadrupole-sextupole-dipole-dipole configuration and a
$\Omega_\pi^{\mathit lab} = 28\,$msr solid angle acceptance each, in
which vertical drift chambers (VDCs) were used for tracking,
scintillation detectors for triggering and timing, and gas {\v
  C}erenkov detectors for discrimination between electrons and
pions. The VDCs are capable of measuring a particle track
%onto either the dispersive or non-dis\-per\-sive plane 
with effective position and angle resolutions of $\sigma_x =
180\,\mu$m and $\sigma_\theta = 1.0\,$mrad.
% for the dispersive focal-plane coordinates.
%Typically the information from 18 drift cells is used to
%determine the track through the chambers leading to a detection
%efficiency very close to 100\,\%. In combination with the very thin
%vacuum windows and the optical design, which corrects for many
%abberrations by hardware, these track detectors allow for 
The spectrometers achieve a relative momentum resolution of $\delta
p/p \sim 10^{-4}$ and were operated at central momenta of 115 and
125\,MeV$\!/c$ with momentum acceptances of $\Delta p/p = 20$\,\%
(SpekA) and 25\,\% (SpekC). The survival probabilities of pions at
these momenta are $\epsilon_\pi \sim 0.3$.
%for flight-paths of $l \sim 10\,$m through SpekA or SpekC. 
%The intrinsic time resolution of the scintillator planes is better
%than $\sigma_t = 255\,$ps after correction for propagation time 
%dispersion.

% ----- kaon tagging ----

The tagging of kaons was performed by the \kaos\ spectrometer.  It was
positioned at zero degrees with respect to the electron beam
direction. The central momentum was 924\,MeV$\!/c$, covering a
momentum range of $\Delta p/p = 50$\,\% with a solid angle acceptance
of $\Omega_K^{\mathit lab} = 16\,$msr.  The detector system includes
segmented scintillator walls for tracking, energy-loss determination
and timing.  Two aerogel {\v C}erenkov detectors were used for pion
rejection with a combined 94\,\% efficiency when keeping the kaon
rejection lower than 1\,\%. The kaon survival probability was
$\epsilon_K \approx 0.40$ for a flight-path of 6.45\,m. The
time-of-flight (TOF) was measured inside the spectrometer with a
resolution of $\sigma_t \approx 180\,$ps along flight-paths of
$1-1.5$\,m. The experimental challenge in this experiment was
originated by the positrons from pair production with large
cross-sections near zero degrees.  The resulting high flux of
background positrons in the spectrometer was reduced by several orders
of magnitude by using a lead absorber with its thickness up to $t =
25\,X_0$ radiation lengths~\cite{Esser2013}. The detection loss for
the kaons in this absorber amounted to $\eta_{\mathit Lead} \sim
70\,$\%.

%%%%%%%%%%%%%%%%%%%%%%%%%%%%%%%%%%%%%%%%%%%%%%%%%%%%%%%%%%%%%%%%%%%%%%%%%%%%%%%
\paragraph{Data analysis---\hspace{-3mm}}
%%%%%%%%%%%%%%%%%%%%%%%%%%%%%%%%%%%%%%%%%%%%%%%%%%%%%%%%%%%%%%%%%%%%%%%%%%%%%%%

The pion momentum, its direction and the reaction vertex were
reconstructed from the focal plane coordinates using the well-known
backward transfer matrices describing the spectrometer optics.  The
momenta of the outgoing pions were corrected for energy-loss inside
the target, a few cm thick of air, and two vacuum window foils
120\,$\mu$m thick each. Kaons were identified by their specific
energy-loss d$E/$d$x$ and velocity $\beta$ from TOF.

% ----- coincidence time spectrum ----

%
\begin{figure}
  \centering
  \includegraphics[angle=0,width=\columnwidth]{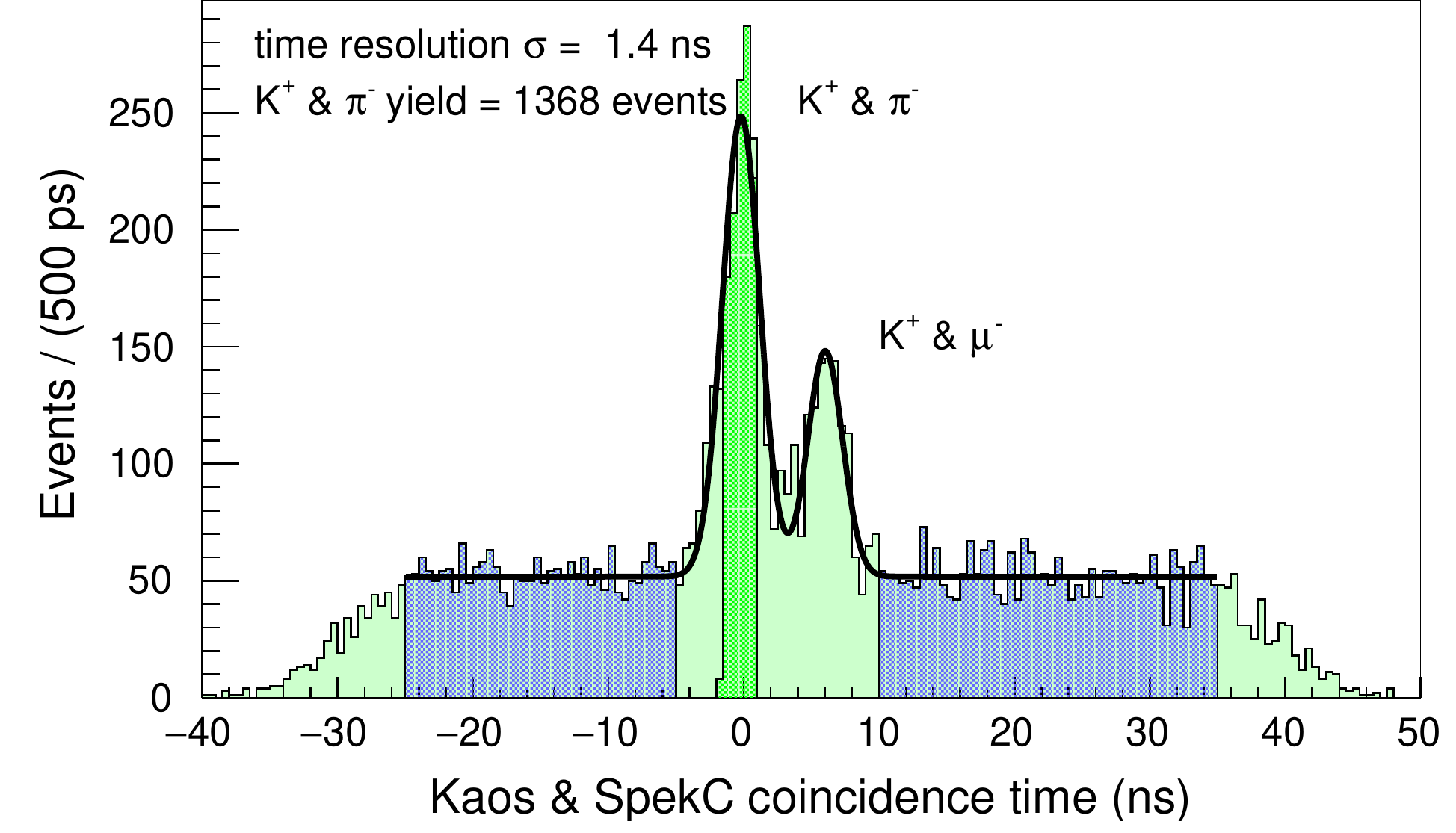}
  \caption{(color online). Coincidence time spectrum for $K^+$ in the
    \kaos\ spectrometer and $\pi^-$ or $\mu^-$ in SpekC, after
    correcting for the reconstructed flight path lenghts of $K^+$ and
    $\pi^-$ through the spectrometers. The time gates for selecting
    true (2.5\,ns width) and accidental coincidences (45\,ns width)
    are indicated by different colors. The solid line represents a fit
    to the spectrum with two Gaussian shaped peaks each on top of a
    accidental background distribution. The peaks were resolved
    with $\sigma_t \sim 1.4$\,ns resolution.}
  \label{fig:SpekC_CoincidenceTime}
\end{figure}

Fig.~\ref{fig:SpekC_CoincidenceTime} shows the coincidence time
between $K^+$ in the \kaos\ spectrometer and $\pi^-$ or $\mu^-$ in
SpekC.  The prominent peak at zero time includes $~10^3$ pions while
the peak of muons is originated by the decay events of pions. True
coincidence events were selected from a time gate with a width of
2.5\,ns. Accidental coincidence events from the two coincidence time
side bands of 45\,ns total width were used to evaluate the accidental
background height and shape in the momentum distribution.

% ----- emulsion data distribution ----

\begin{figure}
  \centering
  \includegraphics[angle=0,width=\columnwidth]{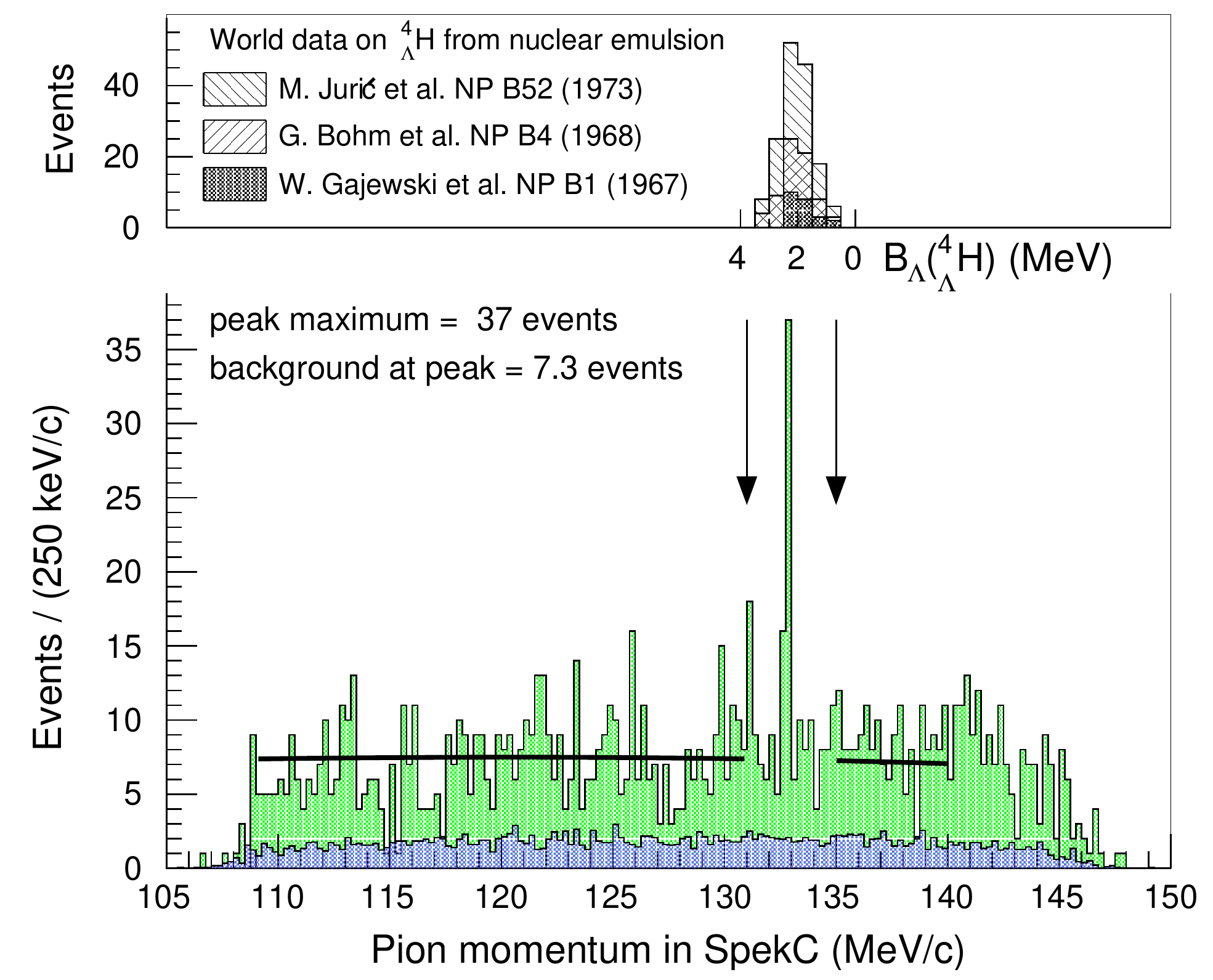}
  \caption{(color online). Pion momentum distribution in SpekC for
    true coincidences (green) and accidental coincidences (blue)
    scaled by the ration of the time gate widths. A monochromatic peak
    at $p_\pi\approx$ 133\,MeV$\!/c$ was observed which is a unique
    signature for the two-body decay of stopped
    $^4_\Lambda\mathrm{H}\rightarrow \mathrm{^4He} + \pi^-$.  The top
    panel shows on the corresponding binding energy scale the
    distribution of data on the $\Lambda$ hyperon binding energy in
    $^{4}_\Lambda$H from emulsion
    experiments~\cite{Gajewski1967,Bohm1968,Juric1973}.  Arrows
    indicate the region of interest in the momentum spectrum.}
  \label{fig:SpekC_PionMomentum}
\end{figure}

The top panel of Fig.~\ref{fig:SpekC_PionMomentum} shows the
distribution of available data on the $\Lambda$ hyperon binding energy
in $^{4}_\Lambda$H from emulsion
experiments~\cite{Gajewski1967,Bohm1968,Juric1973}, where the
compilation in Ref.~\cite{Juric1973} includes re-analyzed events from
Refs.~\cite{Gajewski1967,Bohm1968}. The width of this distribution
from $B_\Lambda > 0.5$\,MeV to $B_\Lambda < 3.5$\,MeV defines a region
of interest corresponding to the momenta of two-body decay pions for
stopped \HH: $131\,\textrm{MeV}\!/c < p_\pi < 135\,\textrm{MeV}\!/c$.

%
% -------------------------------------------------------------------
% -------------------------------------------------------------------
%

% ----- SpekC momentum spectrum ----

The bottom panel of Fig.~\ref{fig:SpekC_PionMomentum} shows the pion
momentum distribution in SpekC for the events within the true
coincidence time gate. The measured pion momentum distribution in the
time side bands was scaled by the ratio of the time gate widths,
giving $1.8 \pm 0.1$ accidental events$/$bin. The exceeding background
was produced by MWD of strange systems, the only reaction that can
generate coincident events meeting the kinematical conditions. The
distribution outside of the region of interest was fitted with a
single scale factor to a template function $bg$ which was determined
by a Monte Carlo simulation of MWD events including angular and energy
dependencies of kaon production in electron scattering off $^9$Be. In
the simulation, the elementary cross-sections for
$p(\gamma,K^+)\Lambda$, $p(\gamma,K^+)\Sigma^0$, and
$n(\gamma,K^+)\Sigma^-$ were taken from the K-Maid
model~\cite{Mart1999,Mart2000} which describes available kaon
photoproduction data. The Fermi-motion effects which modify the
elementary cross-sections for the Be target were calculated in the
incoherent impulse approximation. In the simulated spectrum $\Lambda$
decay-pions are dominating in the range $20-110$\,MeV$\!/c$,
$\Sigma^-$ decay-pions are dominating in the range
$110-194$\,MeV$\!/c$, and at 194.3\,MeV$\!/c$ the monochromatic peak
of stopped $\Sigma^-$ decays is found. Inside the momentum and angular
acceptances of the spectrometer the background spectrum is featureless
and its momentum dependence is practically flat. From the fit result
5.5 events$/$bin can be attributed to $\Sigma^-$ decays in the
measured spectrum.

A localized excess of events over this background was observed inside
the region of interest near to $p_\pi\approx$ 133\,MeV$\!/c$ that is a
unique signature for $^4_\Lambda\mathrm{H}\rightarrow \mathrm{^4He} +
\pi^-$. 
%% In the pion momentum distribution of SpekA no localized
%% excesses of counts over the background expectation of quasi-free
%% hyperon decays were found. 
The region of interest for \HH\ was not inside the acceptance of
SpekA.

% The excess is quantified by its $p$-value for a fluctuation
% in the Poisson distributed background of mean $\lambda = 7.3$ to count
% $n = 37$ or more events: $ p_{local} = 1 -
% \sum_{i=0}^{n-1}\textrm{e}^{-\lambda}\lambda^i/i! = 6 \times 10^{-15}$\,.
% p = probability for counts as extreme or even more extreme as
% actually observed in the localized excess

% A global $p$-value can be computed taking into account the
% probability for such an excess to appear anywhere in the examined
% region of interest: $p = 1 - (1-p_{local})^N$, where $N = 18$ is the
% number of bins in the region. The $p$-values can be converted into
% significance levels $S$ by the association with the normal
% distribution integral probability in which $S$ specifies the number of
% standard deviations through the relation $S = \Phi^{-1}(1-p)$, 
% where $\Phi$ is the inverse of the cumulative distribution function.

% The global p-value is $\sim 10^{-13}$, corresponding to a statistical 
% significance of 7.3.

%%%%%%%%%%%%%%%%%%%%%%%%%%%%%%%%%%%%%%%%%%%%%%%%%%%%%%%%%%%%%%%%%%%%%%%%%%%%%%%
\paragraph{Result and Discussion---\hspace{-3mm}}
%%%%%%%%%%%%%%%%%%%%%%%%%%%%%%%%%%%%%%%%%%%%%%%%%%%%%%%%%%%%%%%%%%%%%%%%%%%%%%%

%
% -------------------------------------------------------------------
\begin{figure}
  \centering
  \includegraphics[angle=0,width=\columnwidth]{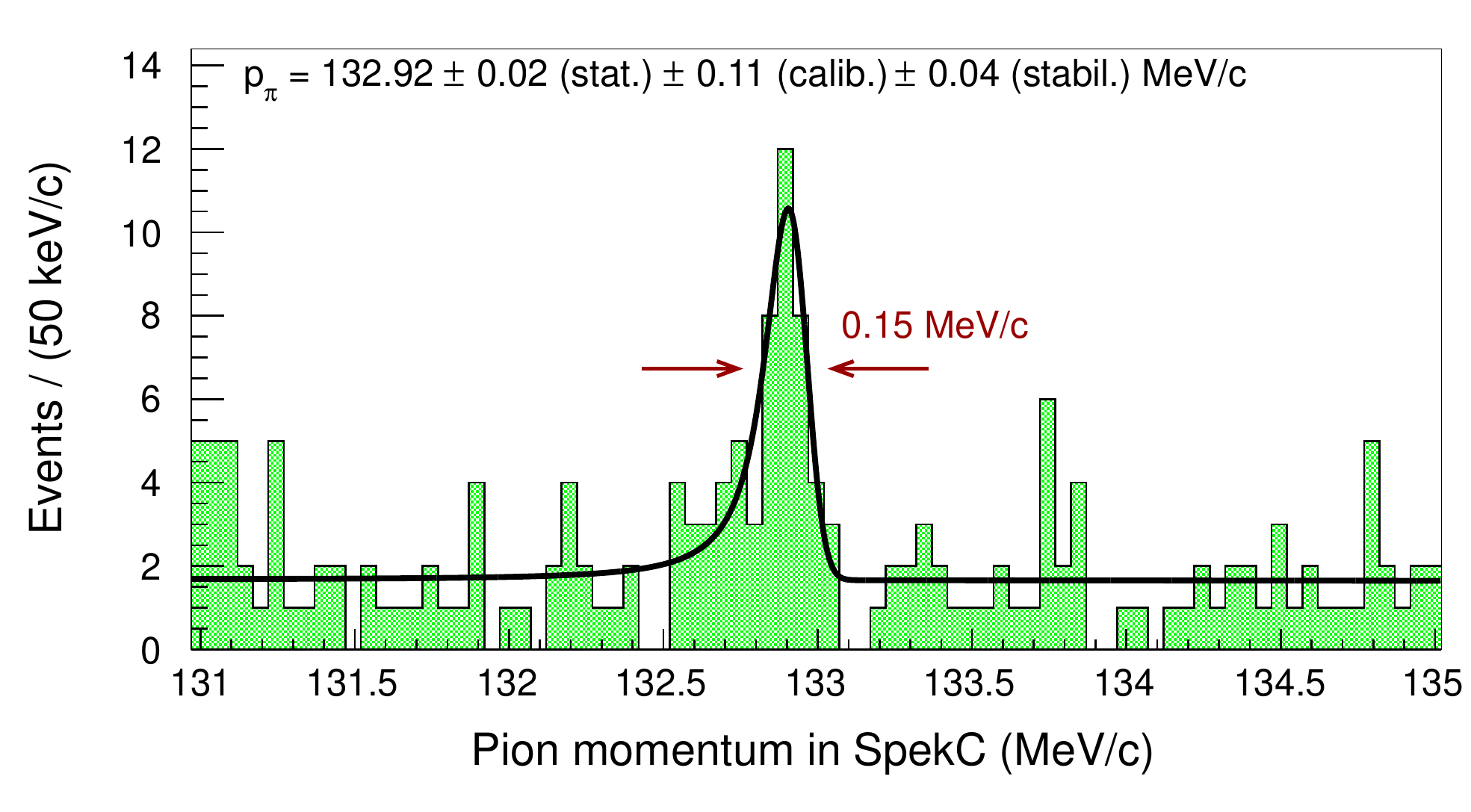}
  \caption{(color online). Pion momentum distribution in SpekC in the
    region of interest with a fit composed of a Gaussian resolution
    function convoluted with a Landau distribution representing the
    energy loss on top of the background function. The observed signal
    shape and width are consistent with the simulation.}
  \label{fig:PeakFit}
\end{figure}
% -------------------------------------------------------------------
%

% ---- fitting of pion momentum spectrum ----

Fig.~\ref{fig:PeakFit} shows the pion momentum distribution in SpekC
in the region of interest. The spectrum was fitted by a function that
is composed of a signal $s$ that is formed by a Landau distribution
representing the known energy loss convoluted with a Gaussian
resolution function on top of the known background $bg$, minimizing
the negative logarithm of the likelihood $L(s+bg)$.  The spectrum was
also fitted with the background-only function.  The corresponding
significance level of the signal calculated via the likelihood ratio
following Ref.~\cite{Cousins2008} is $S_L = \sqrt{-2
  \ln(L(bg)/L(s+bg))} = 5.2$. The shape of the peak was derived from
the simulations of the energy-loss of pions along their tracks with a
most probable energy-loss of $\Delta E \sim 0.140 \pm 0.005$\,MeV. The
width of the peak was observed to be FWHM $\sim 0.15$\,MeV$\!/c$ which
is consistent with the prediction. The largest contribution to the
width was from multiple scattering of the pions inside the target and
at the two vacuum window foils. Uncertainties in the backward transfer
matrix contribute less.

Systematic differences due to the fitting procedure were estimated by
using different probability distribution functions to describe the
peak shape, different fit methods (un-binned and binned) and
minimizers, different fit regions, and different parameterizations of
the background. The most probable momentum for the decay pion peak was
within $\delta p < 10$\,keV$\!/c$ for all cases.

% ---- extraction of lambda binding energy ----

The peak position, $p_{\pi} = 132.92$\,MeV$\!/c$, of the signal was
converted to $\Lambda$ hyperon binding energy, $B_\Lambda$, using
\begin{eqnarray}
  M(_\Lambda^4\mathrm{H}) & = & \nonumber
  \sqrt{M^2(^4\mathrm{He}) + p_{\pi}^2} + \sqrt{M_{\pi}^2 + p_{\pi}^2} 
  \quad \mathrm{and}\\
  B_\Lambda & = & M(^3\mathrm{H}) + M_\Lambda - M(_\Lambda^4\mathrm{H})
  \quad \text{with } c = 1\,,\nonumber
\end{eqnarray}
where the known nuclear masses, $M(\mathrm{^3H}) = 2808.921$\,MeV and
$M(\mathrm{^4He}) = 3727.379$\,MeV, were obtained from tabulated mass
excess values~\cite{Audi2003} and the charged pion mass $M_\pi =
139.570$\,MeV and $\Lambda$ hyperon mass $M_\Lambda = 1115.683$\,MeV
from the latest PDG (Particle Data Group) publication~\cite{PDG2014}.

% ----- assessment of systematic errors ----

The calibration of the momentum spectra has been performed with a
195.17\,MeV electron beam using the $^{181}\mathrm{Ta}(e,e')$ elastic
scattering as well as the inelastic spectrum of the
$^{12}\mathrm{C}(e,e')$ reaction to check the linearity of the
momentum scale.  The momentum was referenced against the beam
energy. The beam energy was measured with an absolute accuracy of
$\delta E_{\mathit beam} = \pm 0.16$\,MeV by exact determination of
the beam position on the accelerator axis and in a higher return
path. The uncertainty on the beam energy translates into a calibration
uncertainty of $\delta p_{\mathit calib.} = \pm
0.11\,\mathrm{MeV}\!/c$. This is the dominant source of systematic
uncertainty in the binding energy. 
% The peak width for the scattered electron momentum was 
% FWHM $= 0.12-0.13\,\mathrm{MeV}\!/c$.
Uncertainties in the spectrometer angle were insignificant. The
backward transfer matrices were checked using sieve slit data.
The stability of the magnetic field in the spectrometers, checked with
regular Hall probe and NMR probe measurements, showed relative
variations of the order $10^{-4}$ that translate into a systematic
uncertainty, $\delta p_{\mathit stabil.}  = \pm
0.04\,\mathrm{MeV}\!/c$, in the momentum.  A total systematic
uncertainty of 0.09\,MeV for the binding energy was obtained using the
kinematical relation $\mathrm{d} B_{\Lambda}/c \approx -
0.725\ \mathrm{d} p_\pi$ for \HH. The final result is then $B_\Lambda
= 2.12 \pm 0.01\ \mathrm{(stat.)} \pm
0.09\ \mathrm{(syst.)}\,\mathrm{MeV}$ with respect to the
$^3\mathrm{H} + \Lambda$ mass.

% ----- discussion of other hypernuclei ----

In the present experiment, the momentum acceptance of the pion
spectrometers covered the monochromatic decay momenta of $^{3,4,6}{}
_\Lambda$H, $^{6,7}{} _\Lambda$He, and $^{7-9}{} _\Lambda$Li,
including very neutron-rich nuclei. A statistical decay model was
applied to evaluate the relative yields~\cite{Esser2013}. From the
range of kinetic energies of the different hypernuclei the stopping
probabilities inside the target were determined: $P_{\it stop} \sim
40\,\%$ for hyperhydrogen isotopes, $70-80\,\%$ for hyperhelium
isotopes and $\sim 90\,\%$ for hyperlithium isotopes. 
% In the combined momentum spectrum above 102\,MeV$\!/c$ no two-body 
% decay peak was observed other than from \HH. One reason is its 
\HH\ has a large total $\pi^-$ decay width, comparable to the free
$\Lambda$ decay width, $\Gamma_{\pi^-}/\Gamma_\Lambda =
1.00\,^{+0.18}_{-0.15}$, and the relative partial decay width of the
two-body mode, $\Gamma_{\pi^- + ^4\mathrm{He}}/\Gamma_{\pi^-} = 0.69
\pm 0.02$~\cite{Outa1998}. These widths are larger than for all
other known light hyperisotopes.

% ----- summary ----

%
% -------------------------------------------------------------------
\begin{figure}
  \centering
  \includegraphics[angle=0,width=\columnwidth]{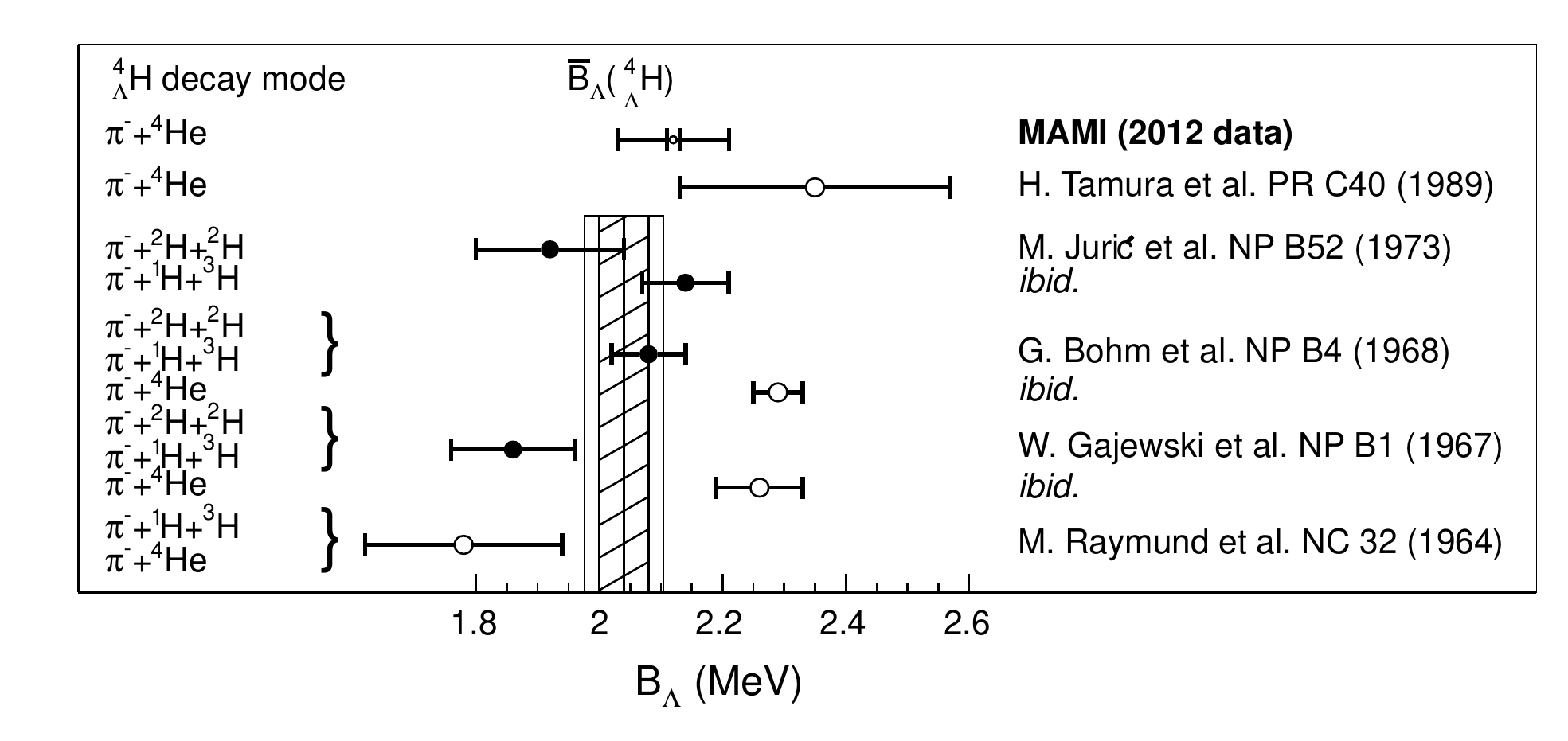}
  \caption{(color online). Measured $\Lambda$ binding energies of
    \HH\ evaluated from pionic
    decays~\cite{Tamura1989,Gajewski1967,Bohm1968,Juric1973,Raymund1964}.
    Full circles present values from three-body decays, open circles
    from two-body decays, errors on the emulsion data are statistical
    only. The binding energy value from Ref.~\cite{Tamura1989} was
    calculated from the observed momentum of $132.6 \pm
    0.3$\,MeV$\!/c$. The mean value as compiled by
    Ref.~\cite{Juric1973} exclude data from the two-body decay mode
    and are shown by the vertical bands with statistical and total
    errors. The uncertainties on the MAMI value are statistical
    (inner) and total (outer).}
  \label{fig:Hyperhydrogen}
\end{figure}
% -------------------------------------------------------------------
%

Detection of the pionic decay of \HH\ has been achieved for the first
time in electroproduction and for the first time using a spectrometer
with $10^{-4}$ relative momentum resolution. This resolution is about
a factor 10 better than for the emulsion
data. Fig.~\ref{fig:Hyperhydrogen} shows a compilation of the
$\Lambda$ binding energy of \HH\ evaluated from pionic decays.  It was
demonstrated that the use of high-resolution spectrometers at a high
intensity beam provides a novel and independent technique to determine
the $\Lambda$ binding energies of light hypernuclei.

%% Fig.~\ref{fig:Hyperhydrogen} shows a compilation of the
%% $\Lambda$ binding energy in the $A = 4$ system evaluated from pionic
%% decays.  The value for $^{4}_\Lambda$He from emulsion data is
%% statistically and systematically better determined than for
%% $^{4}_\Lambda$H. The result from this experiment gives a $\Lambda$
%% binding energy difference $\Delta B_{\Lambda} = 0.27 \pm 0.03\
%% \mathrm{(stat.)} \pm 0.05\ \mathrm{(syst.)} \pm 0.09\
%% \mathrm{(syst.)}$\,MeV, where the first systematic error is the
%% assumed uncertainty for emulsion data and the second systematic error
%% is the uncertainty in the present experiment.  The binding energy
%% splitting in the $A = 4$ system is especially helpful for
%% investigating the origin of CSB.  The new value is a 0.08\,MeV
%% reduction from emulsion data, but still supporting the CSB effect in
%% the system. More constraints will come from measurements of the $1^+$
%% excitation energy by a $\gamma$-ray experiment on $^{4}_\Lambda$He
%% at J-PARC~\cite{JPARC-E13} and the planned reaction spectroscopy of
%% $^4$He$(e,e'K^+)^4_\Lambda$H$(1^+)$ at Jefferson Lab or MAMI.

% ----- acknowledgments ----

\begin{acknowledgments}
This work was supported in part by Deutsche Forschungsgemeinschaft
(SFB 1044), by Carl Zeiss Foundation, by European Community Research
Infrastructure Integrating Activity FP7, by US-DOE contract
DEFG02-97ER41047, by Strategic Young Researchers Overseas Visits
Program for Accelerating Brain Circulation (R2201) and Core-to-Core
program (21002) of JSPS.
\end{acknowledgments}

% ----- references ----

% -------------------------------------------------------------------
%\bibliographystyle{apsrev4-1} 
\bibliographystyle{unsrt} 
\bibliography{references}
% -------------------------------------------------------------------

%%%%%%%%%%%%%%%%%%%%%%%%%%%%%%%%%%%%%%%%%%%%%%%%%%%%%%%%%%%%%%%%%%%%%%%%%%%%%%%
\end{document}